\documentclass{elsart1p}

\usepackage{amssymb}
\usepackage{graphics}
 
\begin{document}
\hfill {\tiny FZJ-IKP-TH-2007-24}
\vspace{-.3cm}

\begin{frontmatter}



\title{$S=-1,-2$ baryon-baryon interactions in chiral effective field theory}


\author{H. Polinder}

\address{Institut f{\"u}r Kernphysik (Theorie), Forschungszentrum J{\"u}lich, D-52425 J{\"u}lich, Germany}

\ead{h.polinder@fz-juelich.de}

\begin{abstract}
% Text of abstract
We have constructed the leading order strangeness $S=-1,-2$ baryon-baryon potential in a chiral effective field theory approach. The chiral potential consists of one-pseudoscalar-meson exchanges and non-derivative four-baryon contact terms. The potential, derived using ${\rm SU(3)}_{\rm f}$ symmetry constraints, contains six independent low-energy coefficients. We have solved a regularized Lippmann-Schwinger equation and achieved a good description of the available scattering data. Furthermore a correctly bound hypertriton has been obtained.

%The chiral potential was used as further input for hypernucleus calculations; a correct hypertriton binding energy and a reasonable $\Lambda$ separation energy for ${}^4_\Lambda{\rm H}$ were found.
\end{abstract}

\begin{keyword}
Hyperon-nucleon interaction \sep Hyperon-hyperon interaction \sep Chiral effective field theory
\PACS 13.75.Ev \sep 12.39.Fe \sep 21.30.-x \sep 21.80.+a
\end{keyword}
\end{frontmatter}

% main text
\section{Introduction}
\label{}

\label{intro}
The derivation of the nuclear force from chiral effective field theory (EFT) has been discussed extensively in the literature since the work of Weinberg \cite{Wei90}. 
An underlying power counting allows to improve calculations systematically by going to higher orders in a perturbative expansion. 
In addition, it is possible to derive two- and corresponding three-nucleon forces as well as external current operators in a consistent way. For reviews we refer the reader to \cite{Bed02,Epelbaum:2005pn}. 
%The main advantages of this scheme are the possibilities to derive two- and three-nucleon forces as well as external current operators in a consistent way and to improve calculations systematically by going to higher orders in the power counting. 
Recently the nucleon-nucleon ($NN$) interaction has been described to a high precision using chiral EFT \cite{Entem:2003ft,Epe05}.
%In \cite{Epe05}, the power counting is applied to the $NN$ potential. The $NN$ potential contains pion-exchanges and a series of contact interactions with an increasing number of derivatives to parameterize the shorter ranged part of the $NN$ force.

As of today, the strangeness $S=-1$ hyperon-nucleon ($YN$) interaction ($Y=\Lambda,\Sigma$) was not investigated extensively using EFT \cite{Sav96}. The strangeness $S=-2$ hyperon-hyperon ($YY$) and cascade-nucleon ($\Xi N$) interactions had not been investigated using chiral EFT so far. In this contribution we show the results for the recently constructed chiral EFT for the $S=-1,-2$ baryon-baryon ($BB$) channels \cite{Polinder:2006zh,Polinder:2007mp}. At leading order (LO) in the power counting, the $YN$, $YY$ and $\Xi N$ potentials consist of four-baryon contact terms without derivatives and of one-pseudoscalar-meson exchanges, analogous to the $NN$ potential of \cite{Epe05}. The potentials are derived using SU(3) constraints. We solve a coupled channels Lippmann-Schwinger (LS) equation for the LO potential and fit to the low-energy $YN$ scattering data. Furthermore results for various $YY$ and $\Xi N$ cross sections are given.

\section{Formalism}
\label{sec:2}
We have constructed the chiral potentials for the $S=-1,-2$ sectors at LO using the Weinberg power counting, see \cite{Polinder:2006zh}. The LO potential consists of four-baryon contact terms without derivatives and of one-pseudoscalar-meson exchanges. The LO ${\rm SU(3)}_{\rm f}$ invariant contact terms for the octet baryon-baryon interactions that are Hermitian and invariant under Lorentz transformations were discussed in \cite{Polinder:2006zh}. The pertinent Lagrangians read
\begin{eqnarray}
{\mathcal L}^1 &=& C^1_i \left<\bar{B}_a\bar{B}_b\left(\Gamma_i B\right)_b\left(\Gamma_i B\right)_a\right>\ , \quad
{\mathcal L}^2 = C^2_i \left<\bar{B}_a\left(\Gamma_i B\right)_a\bar{B}_b\left(\Gamma_i B\right)_b\right>\ , \nonumber \\
{\mathcal L}^3 &=& C^3_i \left<\bar{B}_a\left(\Gamma_i B\right)_a\right>\left<\bar{B}_b\left(\Gamma_i B\right)_b\right>\  .
\label{eq:2.1}
\end{eqnarray}
Here, the labels $a$ and $b$ are the Dirac indices of the particles, the label $i$ denotes the five elements of the Clifford algebra, $B$ is the usual irreducible octet representation of ${\rm SU(3)}_{\rm f}$ (a $3\times3$-matrix). The Clifford algebra elements are here actually diagonal $3\times 3$-matrices in flavor space. The brackets denote taking the trace in flavor space. In LO the Lagrangians give rise to six independent low-energy coefficients (LECs): $C^1_S$, $C^1_T$, $C^2_S$, $C^2_T$, $C^3_S$ and $C^3_T$, where $S$ and $T$ refer to the central and spin-spin parts of the potential respectively.

The contribution of one-pseudoscalar-meson exchanges is discussed extensively in the literature. We do not discuss it here, instead we refer the reader to e.g. \cite{Polinder:2006zh}. 

We solve the LS equation for the $YN$, $YY$ and $\Xi N$ systems. The potentials in the LS equation are cut off with a regulator function, $\exp\left[-\left(p'^4+p^4\right)/\Lambda^4\right]$, in order to remove high-energy components of the baryon and pseudoscalar meson fields. 

\section{Results and discussion}
\label{sec:4}

\begin{figure}[h]
\centering
\resizebox{0.92\textwidth}{!}{%
  \includegraphics*[2.0cm,17.0cm][15.65cm,26.8cm]{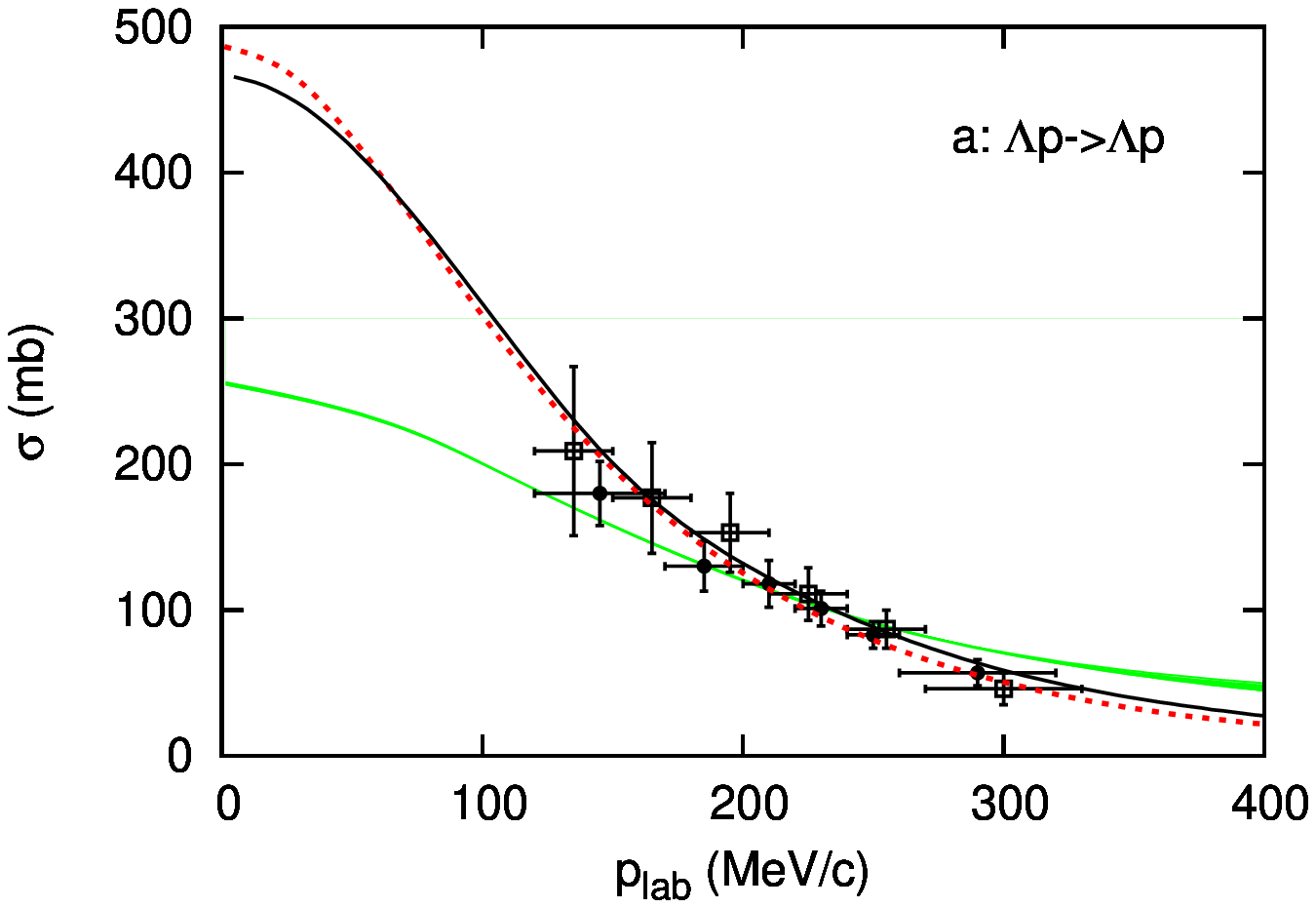}
  \includegraphics*[2.0cm,17.0cm][15.65cm,26.8cm]{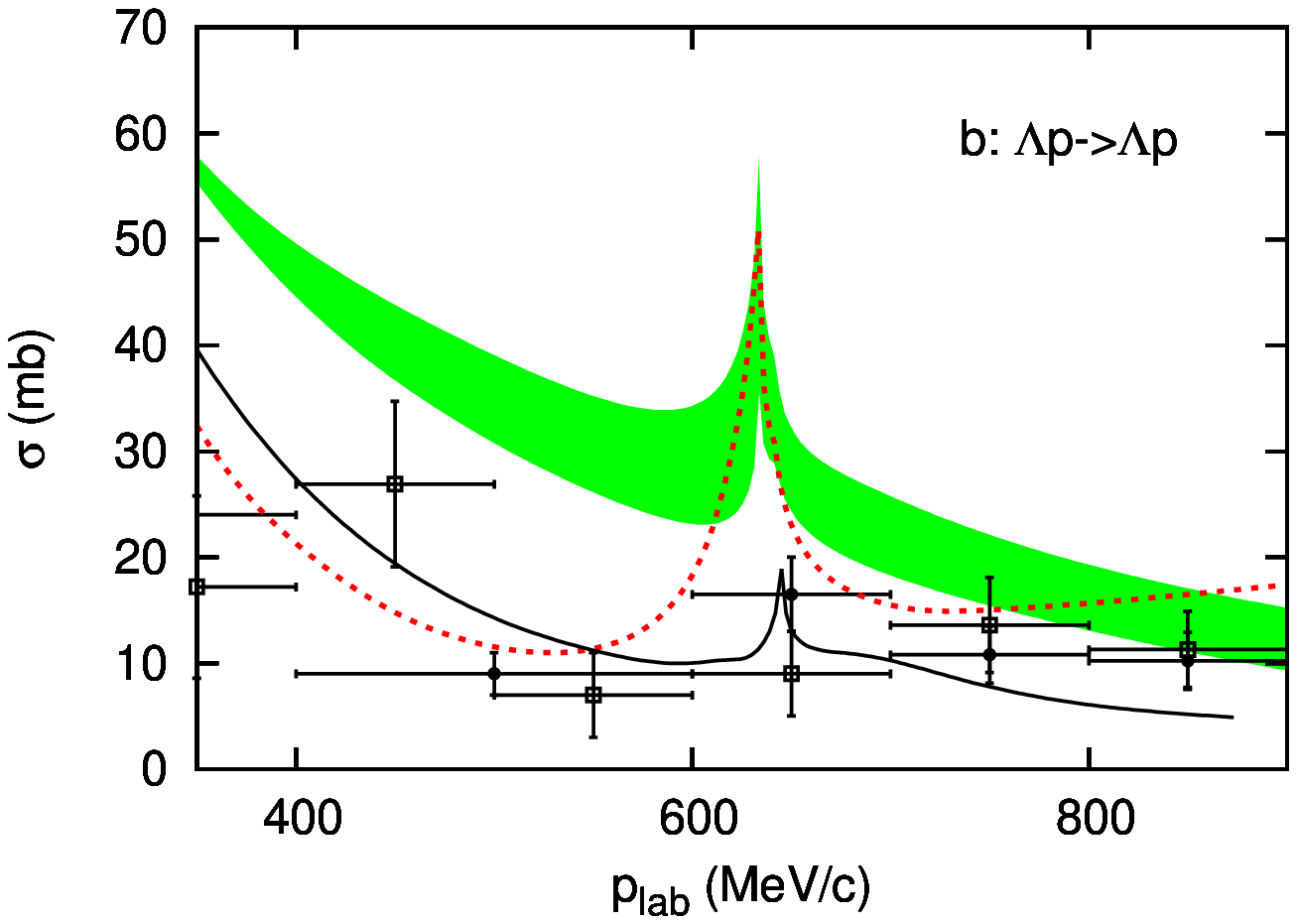}
  \includegraphics*[2.0cm,17.0cm][15.65cm,26.8cm]{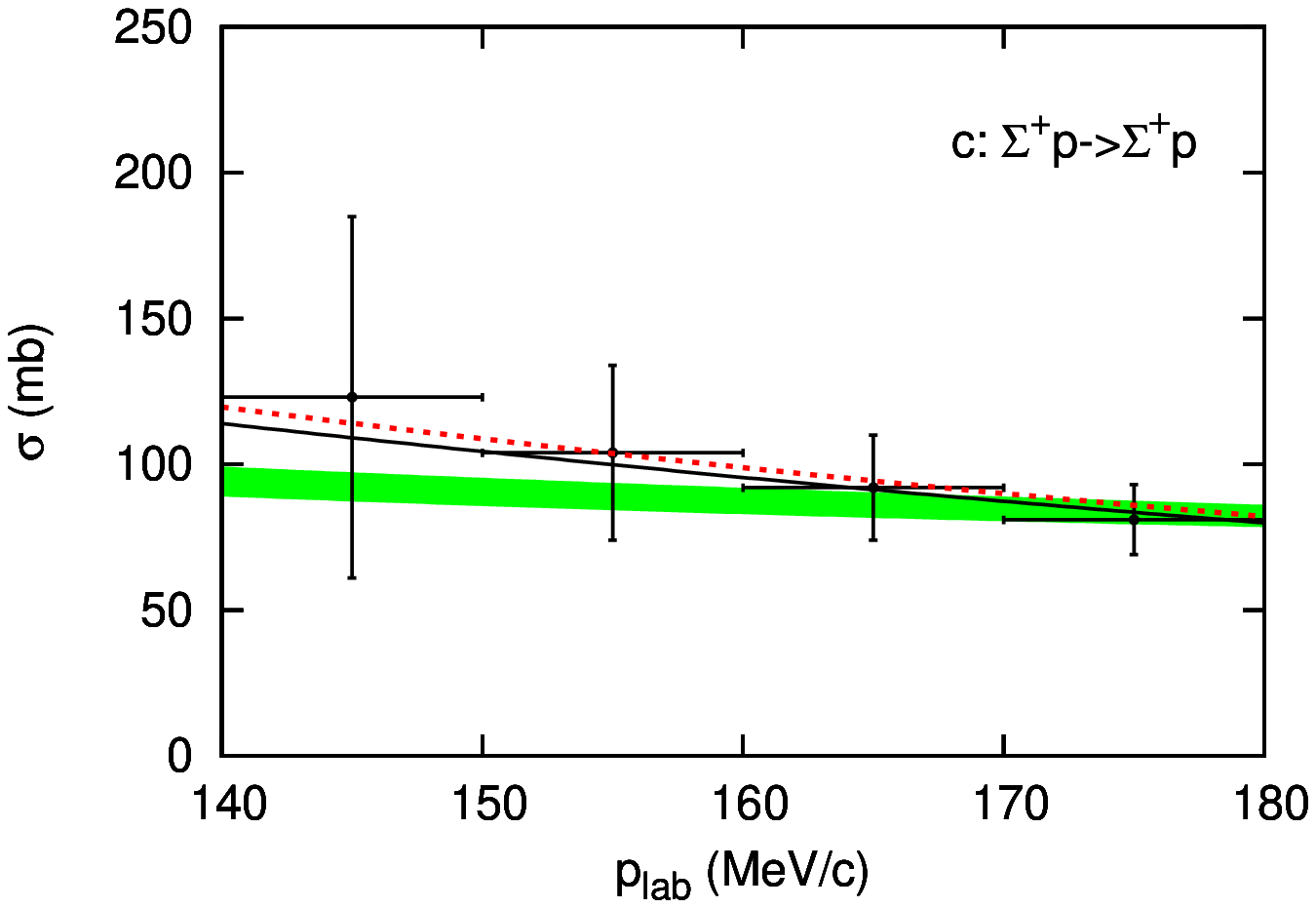}
}
%\hfill \break
\resizebox{0.92\textwidth}{!}{%
  \includegraphics*[2.0cm,17.0cm][15.65cm,26.8cm]{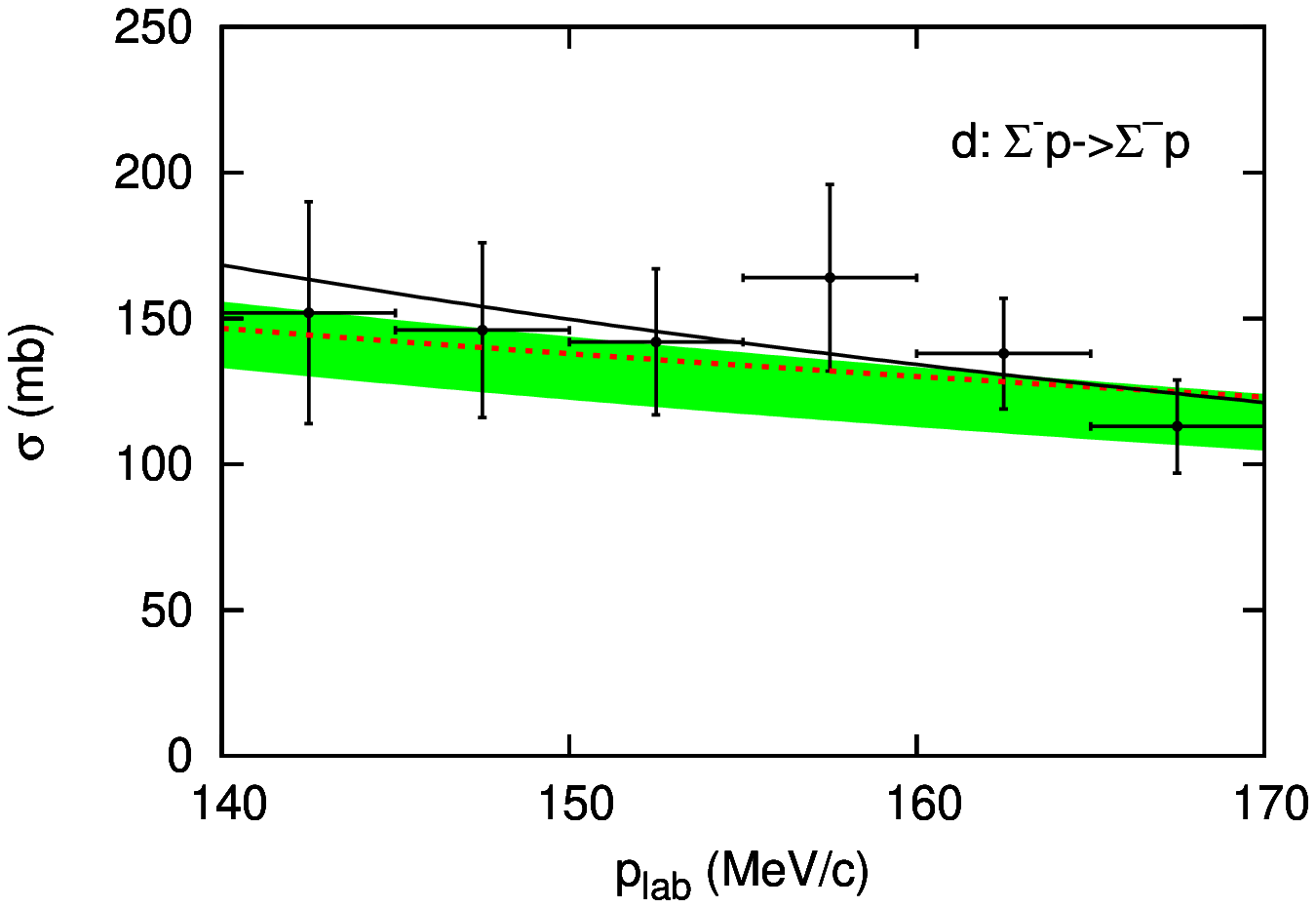}
  \includegraphics*[2.0cm,17.0cm][15.65cm,26.8cm]{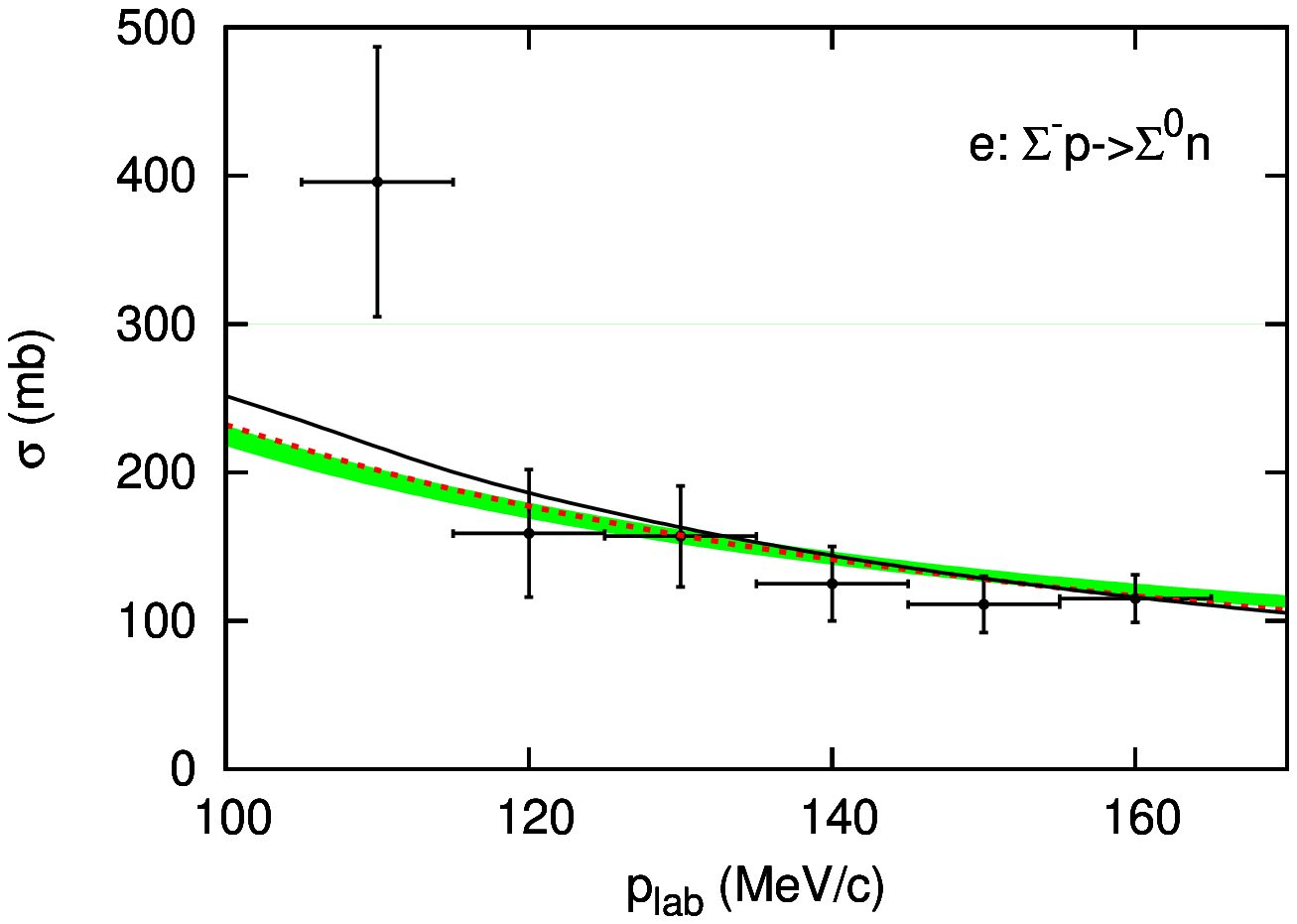}
  \includegraphics*[2.0cm,17.0cm][15.65cm,26.8cm]{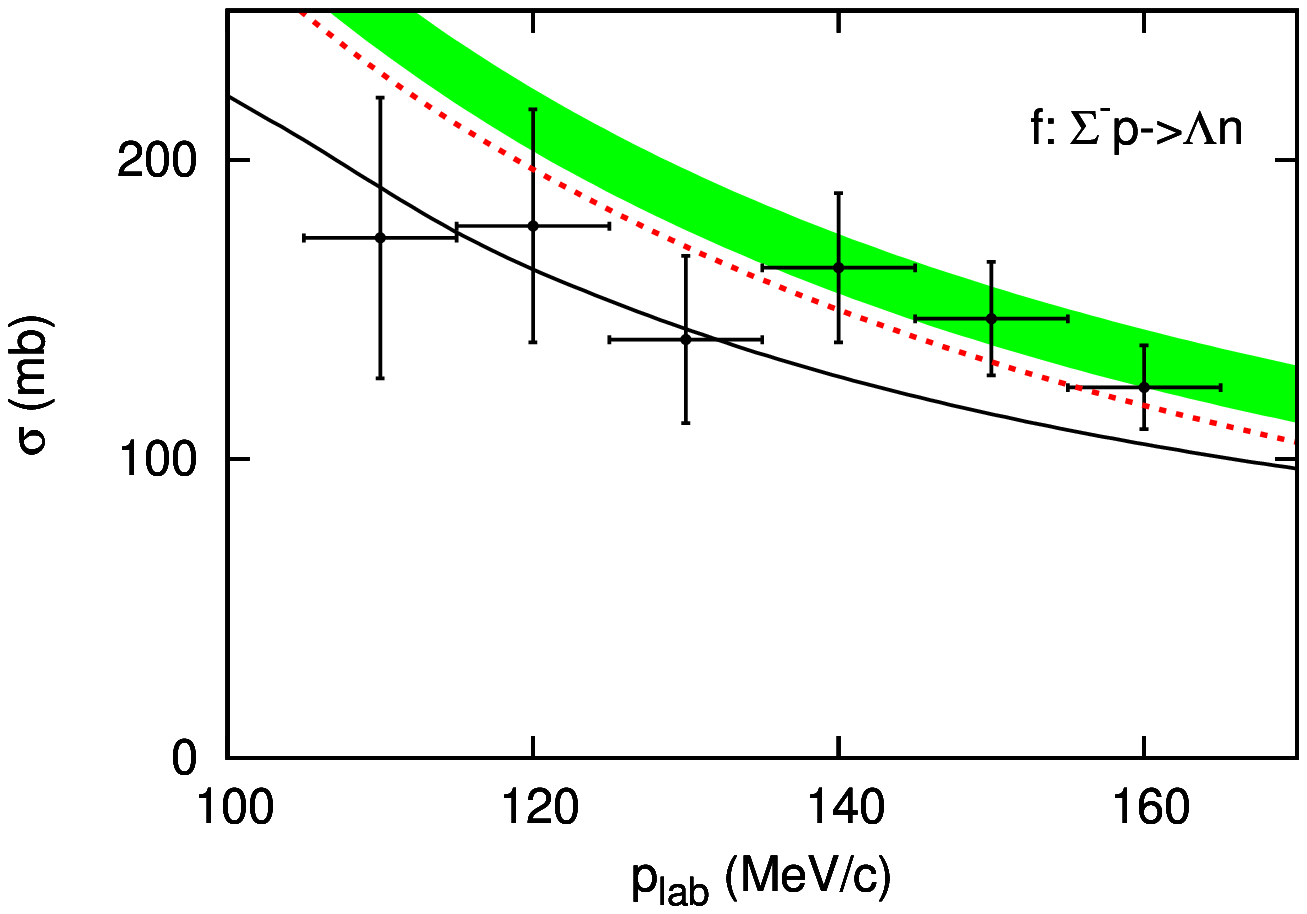}
}
\caption{$YN$ integrated cross section $\sigma$ as a function of $p_{\rm lab}$. The band is the chiral EFT for $\Lambda = 550,...,700$ MeV, the solid and dashed curves are the J{\"u}lich '04 and Nijmegen NSC97f models respectively.}
\label{fig:1}
\end{figure}

\begin{figure}[h]
\centering
\resizebox{0.92\textwidth}{!}{%
  \includegraphics*[2.0cm,17.0cm][15.65cm,26.8cm]{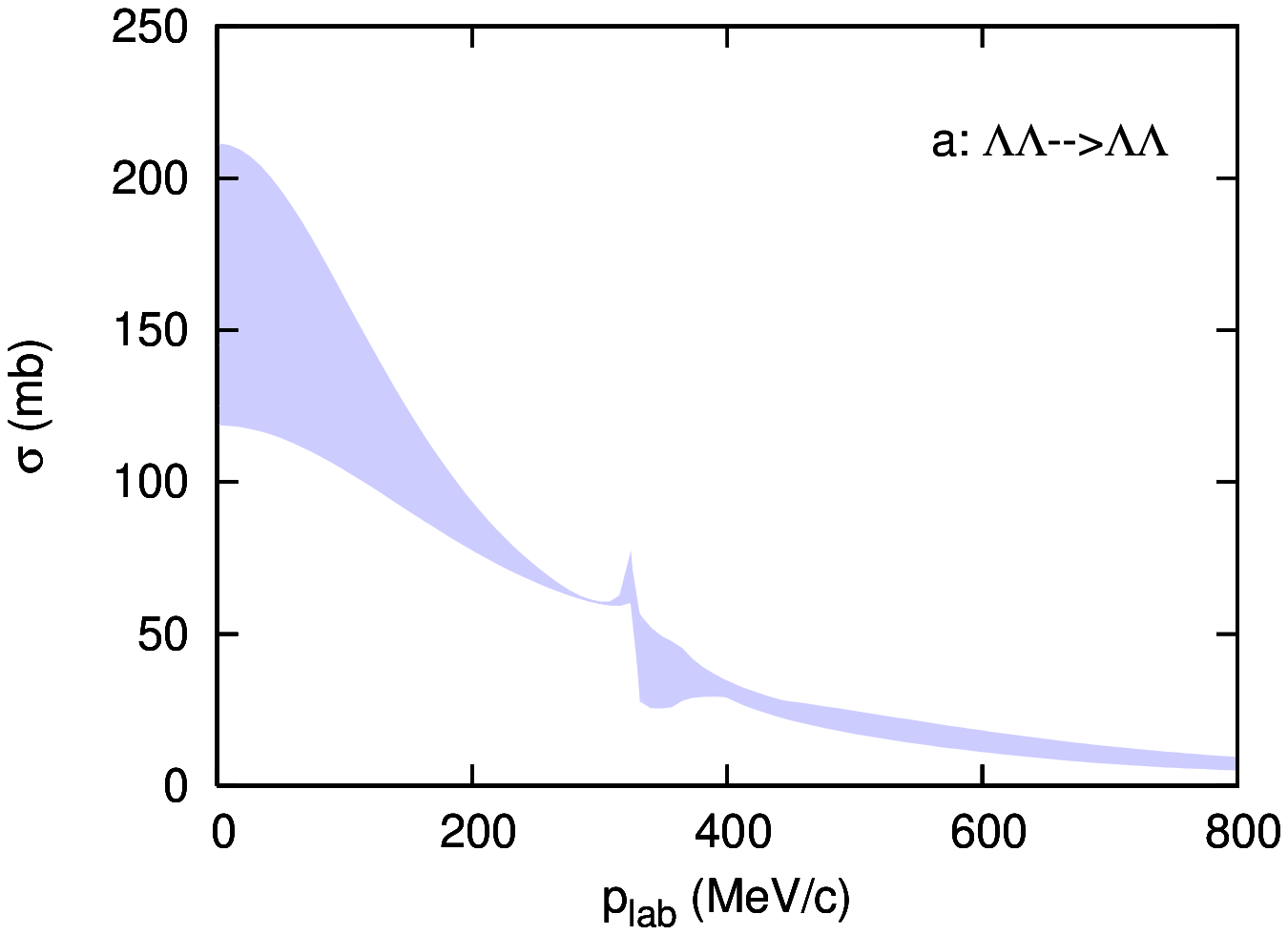}
  \includegraphics*[2.0cm,17.0cm][15.65cm,26.8cm]{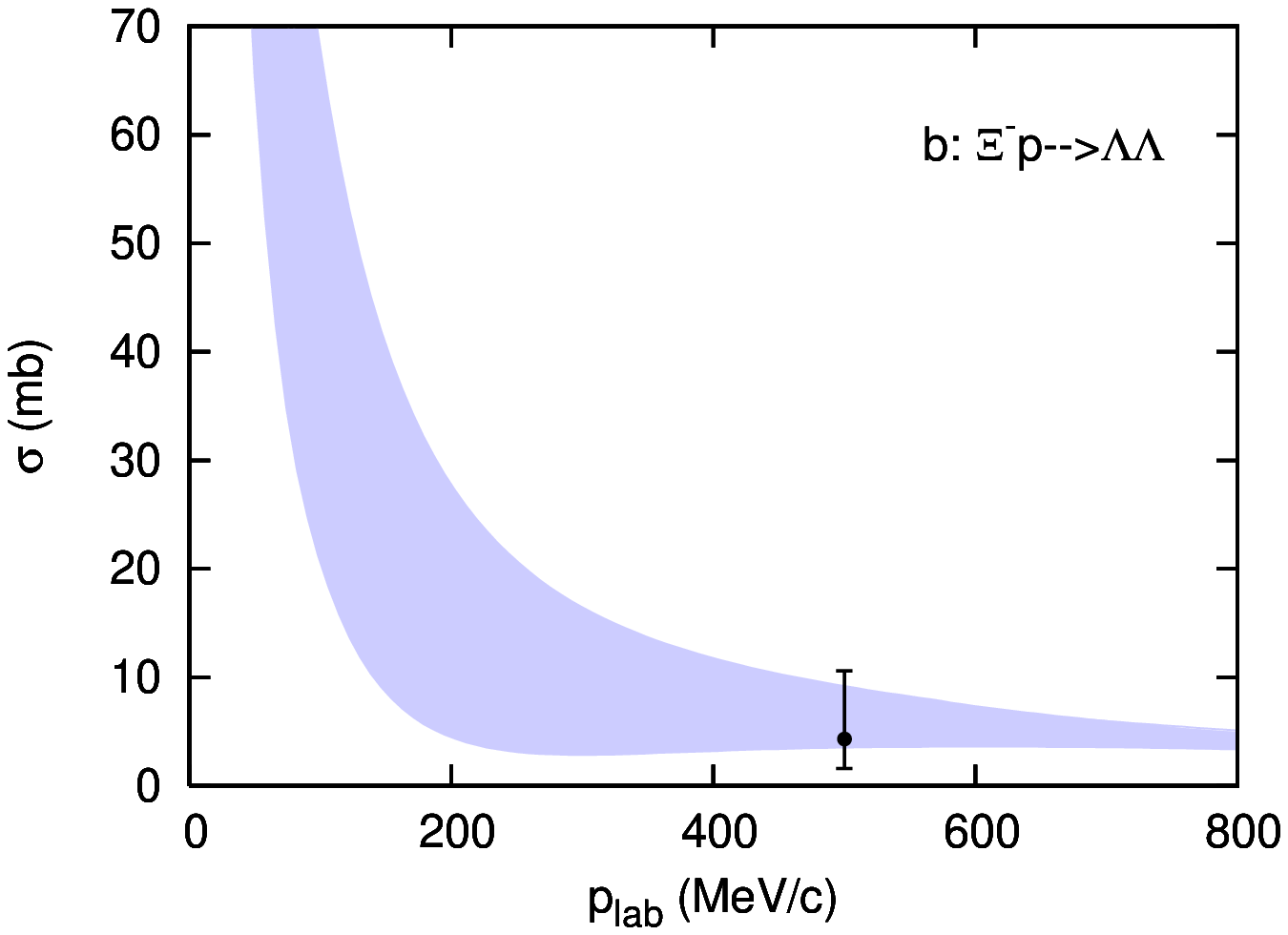}
  \includegraphics*[2.0cm,17.0cm][15.65cm,26.8cm]{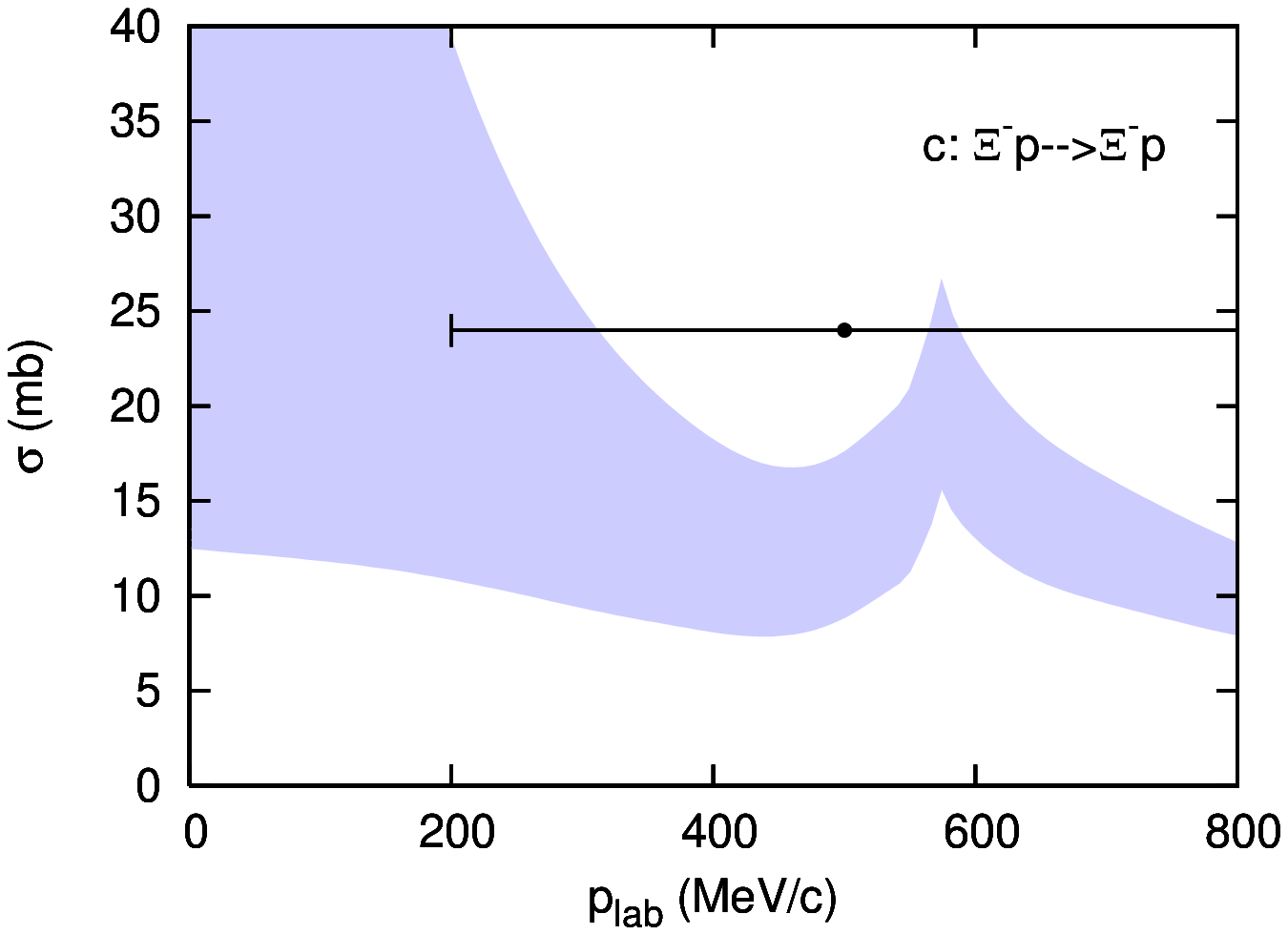}
}
\caption{$YY$ and $\Xi N$ integrated cross section $\sigma$ as a function of $p_{\rm lab}$. The band shows the chiral EFT for variations of the sixth LEC, as discussed in the text.}
\label{fig:2}
\end{figure}

Because of ${\rm SU(3)}_{\rm f}$ symmetry, only five of the LECs can be determined in a fit to the $YN$ scattering data. A good description of the 35 low-energy $YN$ scattering data has been obtained for cut-off values $\Lambda=550,...,700$ MeV and for natural values of the LECs. The results are shown in Fig.~\ref{fig:1}. See \cite{Polinder:2006zh} for more details. In Fig.~\ref{fig:1} the shaded band represents the results of the chiral EFT in the considered cut-off region. For comparison also results for the J{\"u}lich '04 meson-exchange model \cite{Hai05} and the Nijmegen NSC97f meson-exchange model \cite{Rij99} are shown. The $YN$ interaction based on chiral EFT yields a correctly bound hypertriton, also reasonable $\Lambda$ separation energies for ${}^4_\Lambda {\rm H}$ have been predicted \cite{Polinder:2006zh,Nog06a}.

The sixth LEC is only present in the isospin zero $S=-2$ channels. There is scarce experimental knowledge in these channels. In the $\Lambda\Lambda$ system, we assume a moderate attraction and exclude bound states or near-threshold resonances. Based on these considerations the sixth LEC was varied in the range of $2.0,...,-0.05$ times the natural value. Various cross sections for $\Lambda=600$ MeV are shown in Fig. \ref{fig:2}. See \cite{Polinder:2007mp} for more details.

Our findings have shown that the chiral EFT scheme, successfully applied in \cite{Epe05} to the $NN$ interaction, also works well for the $S=-1,-2$ $BB$ interactions in LO. It will be interesting to perform a combined $NN$ and $YN$ study in chiral EFT, starting with a next-to-leading order (NLO) calculation. Work in this direction is in progress.

%\ack
%I thank Johann Haidenbaur and Ulf-G. Mei\ss ner for collaborating on this work. Also I am very grateful to Andreas Nogga for providing me with the hypernuclei results.
%I thank Johann Haidenbaur and Ulf-G. Mei\ss ner for collaborating on this work.

% \appendix

% \section{}
% \label{}

\end{document}